\documentclass[useAMS,usenatbib,amsmath,amssymb]{mn2e}
\usepackage{graphicx}                                                                                                         
\usepackage{wasysym}                                                                                                          
\usepackage{amsfonts}                                                                                                         

\newcommand{\bm}[1]{\mbox{\boldmath $#1$}}  

\title[Tensor Tilt from Primordial B-modes]{Tensor Tilt from Primordial B-modes}
\author[Brian A. Powell]{Brian A. Powell\thanks{E-mail:brian.powell007@gmail.com}\\Institute for Defense Analyses, Alexandria, Virginia, 22311\\}
\begin{document}

\date{\today}

\pagerange{\pageref{firstpage}--\pageref{lastpage}} \pubyear{2011}

\maketitle

\label{firstpage}

\begin{abstract}
A primordial cosmic microwave background B-mode is widely considered a ``smoking gun'' signature of
an early period of inflationary expansion.  However, competing theories of the origin of structure, including string gases and bouncing cosmologies, also produce primordial tensor
perturbations that give rise to a B-mode.  These models can be differentiated by the scale
dependence of their tensor spectra: inflation predicts a red tilt ($n_T<0$),  string gases and loop
quantum cosmology predict a blue tilt ($n_T>0$), while a nonsingular matter bounce gives zero tilt
($n_T=0$).  We perform a
Bayesian analysis to determine how far $|n_T|$ must deviate from zero before a tilt can be detected
with current and future B-mode experiments.  We find that Planck in conjunction with QUIET (II) will
decisively detect $n_T \neq 0$ if $|n_T| > 0.3$, too large to distinguish either single field inflation or string gases from the case $n_T=0$.
While a future mission like CMBPol will offer improvement, only an ideal satellite
mission will be capable of providing sufficient Bayesian evidence to distinguish between each model considered.
\end{abstract}


\section{Introduction}
Precision cosmology is on the verge of answering deep questions about the early Universe.  Are the
temperature fluctuations of the cosmic microwave background (CMB) Gaussian?  Will we discover primordial tensor modes in the CMB?  The existence of a large-scale
spectrum of tensor perturbations is widely considered to be indicative of an early period of inflation -- a ``smoking gun''
signature of quasi-de Sitter expansion;  however, inflation is not the only source
\citep{Brandenberger:2011eq}.
Alternative theories of the origin of structure, like string gas cosmology or
bouncing Universe models, are capable of generating B-modes, as are topological defects
\citep{Seljak:2006hi,Pogosian:2007gi} and bubble collisions
from phase transitions \citep{JonesSmith:2007ne} occurring after the big bang.  These latter two sources have the potential to obscure or confuse the
signal from inflation or its alternatives, but future observations should be able to disentangle the sources if the
primordial signal is large enough \citep{Urrestilla:2008jv,Baumann:2009mq,GarciaBellido:2010if}.  The purpose of the present paper is to investigate how well the different
{\it primordial} sources of tensor modes -- inflation, string gases, or bounces -- can be resolved and
supported. 

The tensor perturbation spectrum is typically modeled as a power law in comoving wavenumber, $k$,
\begin{equation}
P_h(k) = P_h(k_0)\left(\frac{k}{k_0}\right)^{n_T},
\end{equation}
where $n_T$ is the tensor spectral index.  In slow roll inflation the amplitude of tensors is determined by the Hubble
parameter,  $P_h \propto H^2$, which, together with $\dot{H} <
0$ in the Friedmann Universe, furnishes the hallmark prediction of a red spectrum: $n_T < 0$.  Meanwhile, the spectrum
of metric fluctuations that arises
from a string gas in the early Universe is characterized by $n_T > 0$ in the string frame
\citep{Brandenberger:2006xi}, due to the growth of anisotropic
stress as the radiation dominated phase is approached.  A blue spectrum can also be generated
during a phase of
so-called superinflation \citep{Piao:2004tq}, motivated, for example, by loop quantum cosmology \citep{Bojowald:2002nz}.  Yet a different picture emerges from fluctuations produced
during the contracting phase of a bouncing cosmology: a matter dominated contraction with a regular bounce gives a scale invariant spectrum in the expanding radiation dominated phase
\citep{Wands:1998yp,Finelli:2001sr}.  Evidently, the {\it sign}
of $n_T$ is a more likely indicator of inflation than the mere presence of tensors.  But how large must
this tilt be in order for current and future probes to detect it with confidence?   

The basic question that we seek to answer in this work is how far $n_T$ must deviate from zero before one can
confidently conclude $n_T \neq 0$.  From the point of view of sampling
statistics, this is a straightforward hypothesis test: how large must $n_T$ be in order to reject the null hypothesis,
$\mathcal{H}_0: n_T =
0$? Having analyzed data, $d$, one determines whether the best fit value,
$\bar{n}_{T}$, is a sufficient number of standard deviations, $\sigma_{n_T|d}$, away from $n_T = 0$ to satisfy the
chosen significance
level, $\alpha$.  Specifically, assuming that $n_T$ is Gaussian distributed, the p-value must be calculated,
\begin{equation}
\label{pvalue}
p = \frac{1}{\sqrt{2\pi}}\int^\infty_s e^{-z^2/2}\,dz \leq \frac{\alpha}{2},
\end{equation}
where the test statistic $s = |\bar{n}_{T}|/\sigma_{n_T|d}$ is the number of standard deviations that $\bar{n}_{T}$ lies away from $n_T = 0$.
For example, the significance level $\alpha = 0.05$ corresponds to $s = 1.96$, with the conclusion that one risks a 5\%
chance of falsely rejecting the null hypothesis, $n_T = 0$.

If one were to proceed along these lines, it would be determined that the Planck
Surveyor\footnote{http://www.esa.int/planck} will detect $n_T \neq 0$ at
the 95\% confidence level if $|n_T| \geq 0.3$ for a  tensor/scalar ratio $r=0.1$ (see Fig. 1.)\footnote{This result is obtained without the assumption of
Gaussianity leading to Eq. (\ref{pvalue}), but instead follows from a Bayesian parameter estimation performed on a
simulated Planck-precision data set with fiducial tensor spectrum $r=0.1$ and $n_T=0.3$.  We will discuss our
analytical method in detail in the following sections.}  
\begin{figure}
\label{1}
\centering
\includegraphics[width=\linewidth]{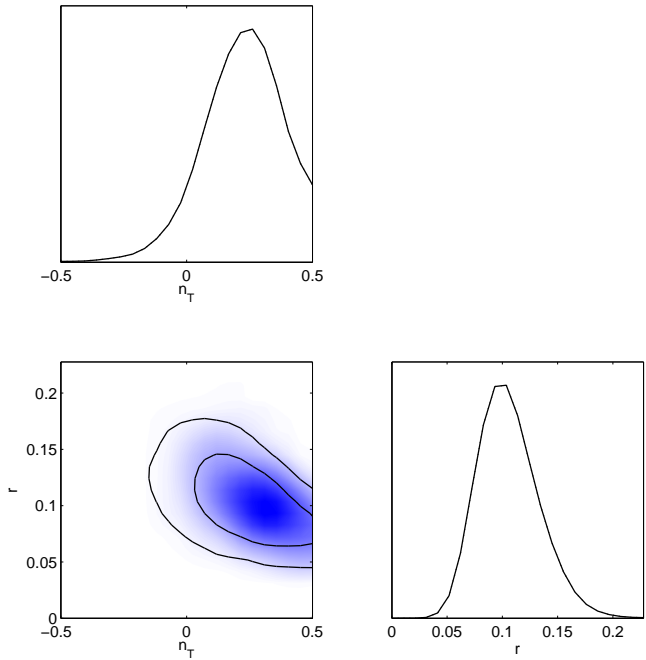}
\caption{A simulated $2\sigma$ detection of $n_T \neq 0$ obtained from mock Planck data with a fiducial tensor spectrum
$r=0.1$ and $n_T =0.3$.}
\end{figure}  
What this approach fails to determine, however, is whether there is a {\it need} for allowing $n_T$ to vary in the
first place: is the fit
to the data sufficiently improved to warrant the inclusion of $n_T$ as a free parameter?  
The accepted approach to this problem is to apply Bayesian inference to the space of competing models. 
As a definition of {\it model}, we should have in
mind a collection of parameters, ${\bm \theta} = (\theta_1,\cdots,\theta_n)$, with associated prior probabilities
$\pi(\theta_1)$, $\cdots$, $\pi(\theta_n)$.  
Here we have two models: $\mathcal{H}_0$ corresponds to the null hypothesis in which $n_T = 0$, and $\mathcal{H}_1$
contains $n_T$ as a parameter free to vary across some prior range,
$[n_{T,{\rm min}},n_{T,{\rm max}}]$.  

Bayesian analysis gives preference to the least complex model that best fits the data
\citep{MacKay,Trotta:2005ar,Liddle:2006tc,Mukherjee:2008zzb,Trotta:2008qt}.  Models with many
parameters free to vary across large prior ranges, $\Delta \theta_j \sim 1/\pi(\theta_j)$, are considered more complex than
models with fewer parameters and more restrictive priors.  Bayesian selection prefers models that are {\it
predictive}: the prior and posterior parameter widths should be comparable.  Frequentist significance tests like Eq. (\ref{pvalue}) do
not incorporate this essential aspect of model selection, and can sometimes give conclusions in striking disagreement with
Bayesian inference \citep{Lindley,Shafer,Trotta:2005ar}.  This disagreement, known as Lindley's
paradox, tends to be most prevalent for detections in the range of two to four $\sigma$ \citep{Trotta:2005ar}: the same
threshold at which the
sampling statistics hypothesis test Eq. (\ref{pvalue}) will detect $n_T = 0.3$ with Planck.  In
this work, we do not solely examine how well future probes might constrain $n_T$, an analysis
carried out in detail by others \citep{Verde:2005ff,Zhao:2009rt,Ma:2010yb};  we
are instead interested in going one step further, and determining whether or not, given these
constraints, we are warranted in including $n_T$ as a free parameter.
In the next section, we
develop the framework for performing Bayesian model comparison forecasts for future experiments, following
closely the approach taken by \citep{Pahud:2006kv} and \citep{Vardanyan:2009ft} to investigate the tilt of
the scalar spectrum and the curvature of the Universe, respectively.

\section[]{Bayesian Evidence}
Methods of Bayesian inference can be applied to the problem of model selection in a manner fully consistent with
probability theory.
Given a model, $\mathcal{H}_i$, and a dataset, {\bf d}, Bayes' Theorem gives
\begin{equation}
\label{evidence}
p(\mathcal{H}_i|{\bf d}) = \frac{p({\bf d}|\mathcal{H}_i)\pi(\mathcal{H}_i)}{\pi({\bf d})},
\end{equation}
where the quantity $p(\mathcal{H}_i|{\bf d})$ is the posterior probability of the model $\mathcal{H}_i$, and $p({\bf
d}|\mathcal{H}_i)$ is the {\it evidence} for model $\mathcal{H}_i$.  One typically assigns equal  prior probabilities $\pi(\mathcal{H}_i)$
to the alternative models, and $\pi({\bf d})$ is independent of the model.  
If we again apply Bayes' theorem,
\begin{equation}
\label{bayes}
p({\bm \theta}|{\bf d},\mathcal{H}_i) = \frac{p({\bf d}|{\bm \theta},\mathcal{H}_i)\pi({\bm
\theta}|\mathcal{H}_i)}{p({\bf d}|\mathcal{H}_i)},
\end{equation}
we obtain the posterior probability of the parameters ${\bm \theta}$ given {\bf d}.  The term $p({\bf d}|{\bm
\theta},\mathcal{H}_i)$ is a function of both the parameter values and the data; for a given dataset, it is commonly
called the likelihood of the parameters, $\mathcal{L}({\bm \theta}|{\bf d},\mathcal{H}_i)$.  The evidence from Eq.
(\ref{evidence}) is nothing more than a normalization constant for Eq. (\ref{bayes}),
\begin{equation}
\label{evidence2}
p({\bf d}|\mathcal{H}_i) = \int \mathcal{L}({\bm \theta}|{\bf d},\mathcal{H}_i)\pi({\bm \theta}|\mathcal{H}_i)d{\bm \theta}.
\end{equation}
This expression defines the evidence  to be the prior-weighted average of the likelihood over the
parameter space.

For the sake of illustration, assume that the parameter $\theta_j$ is well constrained by some data, {\bf d}, so
that the posterior probability $p(\theta_j|{\bf d},\mathcal{H}_i)$ is well-peaked within the prior range,  
\begin{equation} 
\pi(\theta_j|\mathcal{H}_i) = \left[H(\theta_j-\theta_{j,\rm min})H(\theta_{j,{\rm max}}-\theta_j)\right]\frac{1}{\Delta
\theta_j},
\end{equation}
where $H$ is a step function that enforces the prior range and we have chosen a {\it flat} prior on
$\theta_j$, $\pi(\theta_j) = 1/\Delta \theta_j$, within this range.  Since the posterior distribution is well-peaked, we can approximate
the integral Eq. (\ref{evidence2}) using Laplace's method with Eq. (\ref{bayes}): we simply multiply the height of the
un-normalized posterior (the numerator in  Eq. (\ref{bayes})),
$\tilde{p}(\bar{\theta}_{j}|{\bf d},\mathcal{H}_i)$, by its width, $\sigma_{\theta_j|d}$,
\begin{eqnarray}
\label{laplace}
p({\bf d}|\mathcal{H}_i) &=& \int \tilde{p}({\theta_j}|{\bf d},\mathcal{H}_i)d\theta_j,\nonumber\\
&\simeq&  \tilde{p}({\bar{\theta}_{j}}|{\bf d},\mathcal{H}_i) \times \sigma_{\theta_j|d},
\end{eqnarray}
where $\bar{\theta}_{j}$ denotes the point of maximum likelihood.
In terms of the likelihood function this becomes,
\begin{eqnarray}
\label{laplace2}
p({\bf d}|\mathcal{H}_i) &\simeq& \mathcal{L}({\bar{\theta}_{j}}|{\bf d},\mathcal{H}_i)\pi({\theta_j}|\mathcal{H}_i)
\sigma_{\theta_j|d},\nonumber\\
&\simeq& \mathcal{L}({\bar{\theta}_{j}}|{\bf d},\mathcal{H}_i)\frac{\sigma_{\theta_j|d}}{\Delta \theta_j}.
\end{eqnarray}
While only an approximation, this expression nicely reveals the essential ingredients of Bayesian model selection:
a high-valued maximum likelihood clearly increases the evidence in favor of the model, while the {\it Occam factor}, $\beta
= {\sigma_{\theta_j|d}}/{\Delta \theta_j} \leq 1$, penalizes overly complex or poorly predictive models.  Models with
a prior volume much larger than the posterior volume, $\beta \ll 1$, are not considered predictive because they can
accommodate a wide range of parameter values before the data is collected.  Complex models with loose priors and many free
parameters that are well-constrained by the data are therefore penalized by the Occam factor, and will consequently
have lower evidence than a simpler, more predictive model that fits the data equally well.

We are now ready to do model selection: we simply compute the Bayesian evidence for each model and compare.  Given two
competing models, $\mathcal{H}_0$ and $\mathcal{H}_1$, this can be done via the Bayes factor,
\begin{equation}
\label{bf}
B_{01} = \frac{p({\bf d}|\mathcal{H}_0)}{p({\bf d}|\mathcal{H}_1)}.
\end{equation}
A rubric for scoring the significance of a model is given by the well-known Jeffreys' scale \citep{Jeffreys}.
The scale rates: $|\ln B_{01}| < 1$ (indecisive), $1<|\ln B_{01}| < 2.5$
(substantial), $2.5<|\ln B_{01}| < 5$ (strong), and $|\ln B_{01}| > 5$
(decisive), with $\ln B_{01} > 0$  ($\ln B_{01} < 0$) favoring $\mathcal{H}_0$ ($\mathcal{H}_1$).  In this work, we will quote results for strong
and decisive evidence, corresponding to odds ratios of 12:1 and 150:1,
respectively.

While the Laplace method is instructive, it is not sufficient for an accurate determination of the
evidence.  While evaluating the integral in Eq. (\ref{evidence2}) is computationally demanding, various methods have been
applied to problems of model selection in astrophysics, including nested sampling \citep{Skilling,Mukherjee:2005wg} and
thermodynamic integration \citep{Obook,Beltran}.
Here we make use of the {\it Savage-Dickey density ratio} \citep{Dickey}, an exact analytical expression for $B_{01}$ that can be
applied whenever the models to be compared are nested.  Models $\mathcal{H}_0$ and $\mathcal{H}_1$ share the same
cosmological parameters, ${\bm \psi}$, except for $n_T$ which is set to zero in $\mathcal{H}_0$.  For separable
priors, $\pi({\bm \psi}|n_T, \mathcal{H}_1)|_{n_T = 0} = \pi({\bm \psi}|\mathcal{H}_0)$, the Bayes factor takes the form
\begin{equation}
B_{01} = \left.\frac{p(n_T|{\bf d},\mathcal{H}_1)}{\pi(n_T|\mathcal{H}_1)}\right|_{n_T = 0},
\end{equation}
where $p(n_T|{\bf d},\mathcal{H}_1)|_{n_T = 0}$ is the marginalized posterior probability of $n_T$ under model
$\mathcal{H}_1$,
\begin{equation}
p(n_T|{\bf d},\mathcal{H}_1) = \int \mathcal{L}(\{{\bm \psi},n_T\}|{\bf d},\mathcal{H}_1)\,d{\bm \psi},
\end{equation}
evaluated at $n_T = 0$.  This quantity can be obtained relatively easily using Markov Chain Monte Carlo (MCMC) techniques,
and, in principle, gives $B_{01}$ as a function of $n_T$.  This is the function that we seek to determine in this
analysis, for a variety of current and proposed CMB experiments.  For each experiment, we will obtain projections by generating simulated
CMB data across a range of values of $n_T$, and determine how the evidence for $\mathcal{H}_1$ builds as
$|n_T|$ grows.

\section{CMB Spectra from Future Probes}
Primordial perturbations impart inhomogeneities in the
photon temperature at decoupling, measured today as directional anisotropies on the last scattering sphere,
\begin{equation}
\frac{\delta T({\bf n})}{T} = \sum_{\ell,m} a^{T}_{\ell m}Y_{\ell m}({\bf n}),
\end{equation}
where the multipole moments, $a_{\ell m}$, are complex Gaussian random variables with 
variance $\langle a^{T*}_{\ell m}a^T_{\ell' m'} \rangle = C^{TT}_\ell \delta_{\ell
\ell'}\delta_{m m'}$ in the direction {\bf n}.  Quadrupolar temperature anisotropies at decoupling (and again at reionization) are projected into
anisotropies in the polarization of the CMB and can be similarly decomposed \citep{Kamionkowski:1996ks},  
\begin{equation}
\frac{{\mathcal P}_{ab}}{T} = \sum_{\ell m}\left[a^{E}_{\ell m}Y^{E}_{(\ell m)ab}({\bf n})
+a^{B}_{\ell m}Y^{B}_{(\ell m)ab}({\bf n})\right],
\end{equation}
where the $Y^{E,B}_{(\ell m)ab}$ are electric-type (curl-free) and magnetic-type
(divergence-free) tensor spherical harmonics, respectively. The polarization anisotropies are described by the
correlations,
\begin{equation}
\langle |a^{E}_{\ell m}|^2 \rangle = C^{EE}_\ell, \,\,\langle |a^{B}_{\ell m}|^2 \rangle = C^{BB}_\ell,
\end{equation}
and one nonzero cross correlation,
\begin{equation}
\langle a^{T*}_{\ell m}a^E_{\ell m} \rangle = C^{TE}_\ell.
\end{equation}
Primordial density perturbations can be constrained
by measurements of the temperature and E-mode polarization anisotropies, while primordial gravitational waves
additionally create a B-mode polarization pattern \citep{Kamionkowski:1996zd,Seljak:1996gy}.  The B-mode signal is
therefore a key indicator of primordial tensors.

Our projections are based on simulated datasets.  
Since we do not have access to the true distribution of the $a_{\ell m}$'s, an estimator is formed from
their measured  values,
\begin{equation}
\label{est}
\hat{C}^{XY}_\ell = \sum_{m = -\ell}^\ell \frac{|a^{X*}_{\ell m} a^Y_{\ell m}|}{2\ell + 1},
\end{equation}
where $XY = TT$, $EE$, $BB$, and $TE$.   As the sum of the squares of
Gaussian random variables, the $\hat{C}^{XY}_\ell$ are $\chi^2_\nu$-distributed with $\nu = 2\ell + 1$ degrees of freedom. 
We generate simulated data by drawing the $\hat{C}^{XY}_\ell$'s from a $\chi^2_{2\ell + 1}$ distribution with
variances \citep{Knox:1995dq},
\begin{eqnarray}
(\Delta \hat{C}^{XX}_\ell)^2 &=& \frac{2}{(2\ell +1)f_{\rm sky}}\left(C^{XX}_{\ell} + N^{XX}_\ell\right), \\
(\Delta \hat{C}^{TE}_\ell)^2&=& \frac{2}{(2\ell +1)f_{\rm sky}}\left[\left(C^{TE}_{\ell}\right)^2 + \left(C^{TT}_{\ell} + N^{TT}_\ell\right)\right.\nonumber\\
&&\left.\times\left(C^{EE}_{\ell} + N^{EE}_\ell\right)\right],
\end{eqnarray}
where $f_{\rm sky}$ is the fraction of sky covered, the $C^{XY}_\ell$ are the theoretical signal spectra, and
the $N^{XY}_\ell$ are instrumental noise spectra for $X = T$, $E$, and $B$.  For experiments with multiple frequency channels, $c$, the full noise
spectrum is the inverse sum of the individual spectra over the channels,
\begin{equation}
N^{XY}_\ell = \left(\sum_c N^{XY}_{\ell,c}\right)^{-1}.
\end{equation}
Assuming a Gaussian beam, the noise spectrum characterizes the combined effects of the instrument beam smearing and
Gaussian white pixel noise, 
\begin{equation}
\label{noise}
N^{XY}_\ell= (\sigma^{X}_{\rm pix} \theta_{\rm fwhm})^2\exp\left[\ell(\ell+1)\frac{\theta^2_{\rm fwhm}}{8\ln 2}\right]\delta_{XY},
\end{equation}
where $\sigma^{X}_{\rm pix}$ is the noise per pixel (with $\sigma_{\rm pix}^P = \sqrt{2}\sigma_{\rm pix}^T$), $\theta_{\rm fwhm}$ is the full width at half maximum (FWHM) of the
Gaussian beam, and pixel noise from temperature and polarization maps is
uncorrelated and taken to vanish.  

We obtain projections for the Planck Surveyor, both alone and in combination with the ground-based telescopes
PolarBear\footnote{http://bolo.berkeley.edu/polarbear} and the Q/U Imaging Experiment
(QUIET)\footnote{http://quiet.uchicago.edu}, the proposed CMBPol satellite mission \citep{Baumann:2008aj}, and an ideal satellite.  The satellite missions are sensitive to
both the temperature and polarization spectra, while the ground-based experiments will measure only polarization.  The
two platforms are also sensitive to B-modes in different ranges: satellites are most sensitive to the reionization
hump at $\ell < 10$, while ground-based detectors will perform best at intermediate scales, $\ell =30-500$,
probing the anisotropy at decoupling.
In Figure 2 we show the S/N ratios of the experiments considered in this analysis as a function of
multipole number, $\ell$, for a B-mode signal with $r = 0.1$.

For Planck, we include three channels with frequencies (100 GHz, 143 GHz, 217 GHz) and noise levels per beam
$(\sigma^{T}_{\rm pix})^2 =$ ($46.25$ $\mu$K$^2$, $36$ $\mu$K$^2$, $17.6$ $\mu$K$^2$). The FWHM of the three channels are $\theta_{\rm fwhm} =$ (9.5', 7.1', 5.0') \citep{planck:2006uk}. 
For PolarBear we consider frequency channels (90 GHz, 150 GHz, 220 GHz) with $(\sigma^{T}_{\rm pix})^2 =$ ($2.6$
$\mu$K$^2$, $5.8$ $\mu$K$^2$, $744$ $\mu$K$^2$) and resolutions $\theta_{\rm fwhm} =$ (6.7', 4.0',
2.7'), and for
QUIET (II) we use frequencies (60 GHz, 90 GHz) with $(\sigma^{T}_{\rm pix})^2 =$ ($0.04$ $\mu$K$^2$, $0.08$
$\mu$K$^2$) and $\theta_{\rm fwhm} =$ (23', 10') \citep{Samtleben:2008rb}.  We assume integration times of 0.5 and 3 years for these two experiments, respectively.
We combine three channels for CMBPol with frequencies (100 GHz, 150 GHz, 220 GHz) and noise levels
$(\sigma^{T}_{\rm pix})^2 =$($729$ nK$^2$, $676$ nK$^2$, $1600$ nK$^2$) and $\theta_{\rm fwhm} =$ (8', 5', 3.5')
\citep{Fraisse:2008ar}.
In Figure 2, for reference, we also include the S/N ratio with lensing as the only noise source (green dashed).  This contamination, arising from the gravitational
lensing by large scale structure of primordial E-modes into B-modes \citep{Zaldarriaga:1998ar}, dominates the signal at $\ell > 100$; probes that seek to measure
the B-mode signal in this range must successfully subtract the lensing contribution.  This effect helps motivate our
selection of experiments: PolarBear's sensitivity lies just below the lensing signal, while the full capability of the
upgraded QUIET (II) and CMBPol will be more sensitive and require successful delensing
\citep{Hu:2001kj,Hirata:2003ka}.  We assume for these experiments that
lensing is reduced to a level of $5\times 10^{-8}$ $\mu{\rm
K}^2$ \citep{Seljak:2003pn}.  Lastly, our ideal experiment will suffer from zero
instrumental noise or beam effects, but will still be limited by cosmic variance, $(\Delta
\hat{C}_\ell)^2 = 2C_\ell^2/(2\ell + 1)$, and the reduced
lensing noise.   We assume sky coverages of $f_{\rm sky} =$ 0.65, 0.012, 0.04, 0.8, and 1.0,
respectively, for the experiments just listed.  We next present our analysis and results.
\begin{figure}
\label{2}
\centering
\includegraphics[width=3.25in]{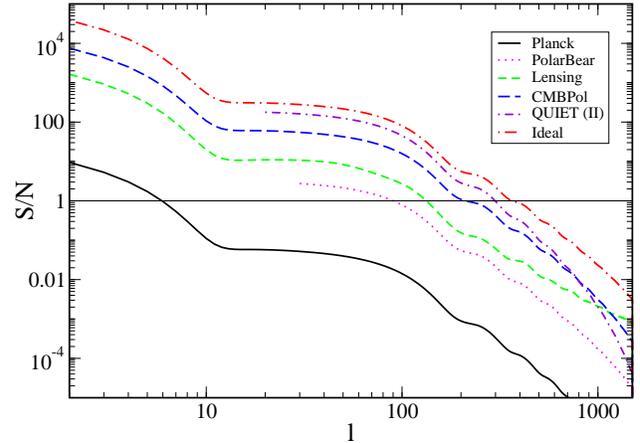}
\caption{Signal-to-noise ratios of experiments considered in this analysis as a function of multipole
number,
$\ell$, based on a fiducial model with $r=0.1$.}
\end{figure}  

\section{Evidence for a Tensor Tilt}
The determination of how $B_{01}$ varies with $n_T$ requires an evaluation of $p(n_T|{\bf d},\mathcal{H}_1)$ (c.f. Eq. \ref{bf}) across a
range of $n_T$, although, in general, this distribution also depends implicitly on $r$ ($\sigma_{n_T}$ is
an increasing function of $r$.)  We first confront the
simpler task of obtaining results for
$B_{01}$ with the fiducial tensor amplitude fixed at $r=0.1$, near the upper 95\% C.L. set by
WMAP+BAO+$H_0$ \citep{Komatsu:2010fb}.
Though limited, this provides an optimistic projection and filters out those experiments that fail to provide strong evidence for
$n_T\neq 0$ even under the most favorable conditions; those experiments that do provide strong evidence are 
analyzed further for the case of variable $r$.
\subsection{Optimistic Projection}
For each experiment listed in the previous section, we obtain $p(n_T|{\bf d},\mathcal{H}_1)$ by
analyzing simulated
data generated from the model $\mathcal{H}_1$.  We assign the base cosmological parameters the fiducial values $\Omega_b
h^2 = 0.022$, $\Omega_c h^2 = 0.105$, $\theta_s = 1.04$, $\tau = 0.09$, $A_s = 2.23\times 10^{-9}$, $n_s = 0.97$, $r = 0.1$, and
simulate a different dataset as $n_T$ is incremented in steps of $|\delta_{n_T}|= 0.1$ across the prior range $n_T \in
[-0.5,0.5]$\footnote{The Bayes factor depends on the prior range, but only weakly:  for example, doubling the
range gives $|\Delta \ln B_{01}| = \ln 2 = 0.7$.}.  We use MCMC to constrain $r$ and $n_T$ for each dataset; since the base
parameters $\Omega_bh^2$, $\Omega_c h^2$, $\theta_s$, $\tau$, $A_s$, and $n_s$ have little effect on the B-mode signal, we
only vary $r$ and $n_T$ within the chains. 
The theoretical temperature and polarization $C_\ell$-spectra are generated out to $\ell=2000$ with
\texttt{CAMB}\footnote{http://camb.info}
\citep{Lewis:1999bs}, and the
parameter estimation is performed using \texttt{CosmoMC}\footnote{http://cosmologist.info/cosmomc} \citep{Lewis:2002ah}.  For each dataset, convergence is measured across four chains using the
Gelman-Rubin R statistic.
The constraints on $r$ and $n_T$ are in general correlated, but the parameter uncertainties can be minimized by choosing
the pivot scale corresponding to the multipole at which they become uncorrelated, $k_* \simeq 10^{-4}{\rm
Mpc}^{-1}\ell_*$.  This pivot scale depends on the data, and will be different for
each experiment: $\ell_* = 10, 30, 65, 150,$ and 300 for Planck, Planck+PolarBear,
Planck+QUIET (II), CMBPol, and the ideal experiment \citep{Zhao:2009mj,Zhao:2009rt}. By constraining
$r(k_*)$, we minimize the effects of correlations in our comparisons of constraints across experiments.

We present results in Figure 3 and Table 1.  
In Figure 3, the data points represent the actual evaluations of $\ln B_{01}$; the curves are obtained
via quadratic regression.  The gray shaded region, $n_T > 0.15$, is ruled out by nucleosynthesis constraints on the
energy density of gravitational waves with power law spectra \citep{Stewart:2007fu}.  
We find that Planck by itself will not find decisive (strong)
evidence for $n_T \neq 0$ unless $|n_T| > 0.5$ (0.43); in combination with PolarBear and QUIET (II), these same levels
of evidence are achieved
for $|n_T| \approx 0.36$ (0.28) and $0.29$ (0.25), respectively.  These detection thresholds are ruled out for power law
spectra.  The constraint from nucleosynthesis should, however, be applied with care since the assumption of a power law tensor spectrum
is not always appropriate\footnote{While power law tensor spectra are expected from slow roll inflation, the primordial gravitational waves generated in
scenarios of loop quantum cosmology are strongly non-power law: while typically steeply blue ($n_T \approx 2$) on CMB
scales, they are
scale invariant on smaller scales.  We discuss this further in Section 5.}.  Meanwhile, a future space-based mission with the
specifications of CMBPol will provide decisive evidence of $n_T \neq 0$
for $|n_T| \geq 0.15$ -- just on the edge of the excluded region, and strong evidence for $|n_T| \geq 0.12$.  Lastly, an ideal satellite experiment will perform even better: not only will it support
a decisive (strong) confirmation of $n_T \neq 0$ for $|n_T| > 0.03$ (0.025), but it will provide strong evidence for $n_T = 0$, {\it i.e.} favor the null
hypothesis $\mathcal{H}_0$, if $|n_T| < 0.01$.  These results are summarized in Table 1.

\begin{figure}
\centering
\includegraphics[width=3.25in]{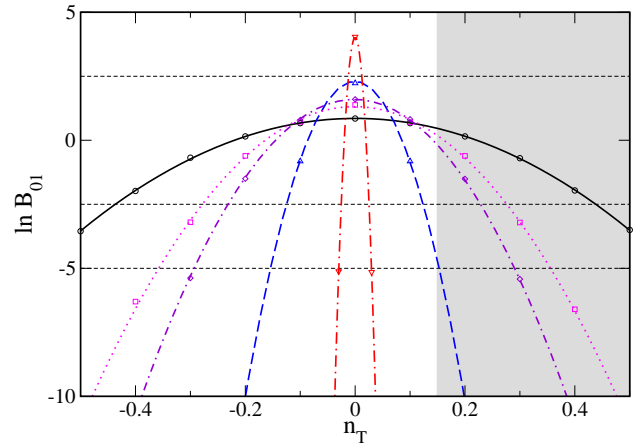}
\caption{Bayes factor as a function of $n_T$ with $r=0.1$ for the experiments discussed in the text: Planck (black solid), Planck+PolarBear (magenta
dotted), Planck+QUIET (II) (purple dash dot), CMBPol (blue dashed), and ideal satellite (red long dash dot).  The gray region is excluded
for power laws by nucleosynthesis constraints.}                                                                              
\end{figure}  
\begin{table}                                                                                                                
\caption{Fiducial $n_T$ for which the different experiments will support strong ($|\ln B_{01}| = 2.5$) and decisive
($|\ln B_{01}| = 5$) evidence in favor of $n_T \neq 0$ ($\ln B_{01} < 0$) and $n_T=0$ ($\ln B_{01} >0$).} 
\begin{tabular}{l|c|c|c}                                                                                                     
&$|n_T|$ & $|n_T|$&$|n_T|$  \\
Experiment&$\ln B_{01} = -2.5$ &$\ln B_{01}= -5$&$\ln B_{01} = 2.5$\\\hline
Planck&  0.43 & $>0.5$&- \\
\hspace{2mm}+PolarBear& 0.28 & 0.36 &-\\
\hspace{2mm}+QUIET (II)& 0.23 & 0.29&- \\
CMBPol& 0.12 & 0.15&- \\
Ideal& 0.025 & 0.03&0.01 \\         
\end{tabular}                                                                                                                
\end{table}
\subsection{Dependence on $r$}
The previous findings apply to $\mathcal{H}_1$ with fiducial $r=0.1$.  However, the error on $n_T$, which largely determines the Bayes
factor, depends on the base value of $r$.  We now examine whether and how our conclusions change when $r$ is also
allowed to vary.  The preferred approach to this problem would be apply the same MCMC analysis on data
sets generated from a range of fiducial $r$, in addition to $n_T$.  This method, however, is time
consuming and more efficient approaches exist. The previous results from the MCMC analysis show 
that the posterior distributions of $r(k_*)$ and $n_T$ are nearly Gaussian and not too strongly correlated, and so
a Fisher matrix analysis should provide reliable constraints \citep{Perotto:2006rj}. 

A forecast of parameter constraints can be obtained relatively easily by Taylor expanding the log-likelihood
function, $\ln \mathcal{L}(\bm \theta | \bf d)$, about the best-fit parameter values, $\bar{\bm \theta}$, and examining
the 2$^{nd}$-order coefficient,
\begin{equation}
\label{fish}
F_{ij} = \left.-\frac{\partial^2 \ln \mathcal{L}}{\partial \theta_i \partial \theta_j}\right|_{\bm \theta = \bar{\bm
\theta}}.
\end{equation}
The Fisher information matrix, $F_{ij}$, encodes parameter correlations and measures the steepness of the
likelihood function in the direction of each parameter $\theta_i$.  The minimum precision with which parameter
$\theta_i$ can be measured is set by the Cramer-Rao bound \citep{Tegmark:1996bz},
\begin{equation}
\label{cr}
\sigma_{\theta_i} \geq \sqrt{(F^{-1})_{ii}}.
\end{equation} 
We again consider only the parameters $r$ and $n_T$, and only include B-mode data in the likelihood
function \citep{Bond:1998qg},
\begin{eqnarray}
\label{blike}
-2\ln \mathcal{L} &=& \sum_\ell (2\ell + 1)f_{\rm sky}\left\{\ln\left(\frac{C^{BB}_\ell + N^{BB}_\ell}{\hat{C}^{BB}_\ell}\right)\right.\nonumber\\
&& \left.+ \frac{\hat{C}^{BB}_\ell}{C^{BB}_\ell + N^{BB}_\ell} - 1\right\},
\end{eqnarray}
where $\hat{C}^{BB}_\ell$ and $N^{BB}_\ell$ are defined in Eqs. (\ref{est}) and (\ref{noise}), respectively, and
$C^{BB}_\ell$ is the theoretical spectrum.  
Using Eq. (\ref{blike}) in Eq. (\ref{fish}) gives
\begin{equation}
\label{fishcl}
F_{ij} =  \frac{1}{2}\sum_\ell \left(2\ell + 1\right)f_{\rm sky}\frac{\partial C^{BB}_\ell}{\partial
\theta_i}{\left(C^{BB}_\ell + N^{BB}_\ell\right)^{-2}}\frac{\partial C^{BB}_\ell}{\partial \theta_j}.
\end{equation}

We now extend our MCMC
analysis of CMBPol and the ideal satellite experiment to examine how these projections vary with fiducial $r$ (Planck
and the ground-based experiments fail to provide strong evidence for $n_T \neq 0$ consistent with nucleosynthesis
constraints even in the optimistic case.)  We compute
the Fisher matrix Eq. (\ref{fishcl}) for each experiment on a 200$\times$200 grid in the $(n_T,\log r)$ plane with $r \in [10^{-6},0.4]$ and
$n_T \in [-0.5,0.5]$. Having obtained $\sigma_r$ and $\sigma_{n_T}$ from Eq. (\ref{cr}), we obtain the Bayes factor via
Eq. (\ref{bayes}) under the assumption that the marginalized posterior distributions of $r$ and $n_T$ are Gaussian.  
We
present results for CMBPol and the ideal satellite experiment in Figure 4.  The CMBPol thresholds for strong and
decisive evidence are indicated in violet and blue, respectively, and the same thresholds for the ideal
experiment are given in orange and red, respectively.  
For power law spectra, CMBPol will find strong evidence for
$\mathcal{H}_1$ if $r \geq 0.06$, and decisive evidence if $r \geq 0.1$ (in this case confirming the optimistic results of the MCMC
analysis, Figure 3.)  The ideal experiment has a much better chance of detecting $n_T \neq 0$: $r \geq 10^{-3}$ and $r
\geq 10^{-4}$ for decisive and strong evidence, respectively.  The ideal satellite will also find strong evidence in
favor of the null hypothesis, $\mathcal{H}_0$, for $r>2.5\times 10^{-3}$ and $n_T$ within the orange
dotted contour in Figure 4.  

\section{Discussion}
We have obtained projections for the detectability of $n_T\neq0$ with current and future satellite
and ground-based CMB experiments.  Current and proposed missions will struggle to detect a tensor
tilt with statistical significance; however, an ideal satellite experiment will perform well.
This indicates that discriminating between theories of early Universe structure formation with
primordial B-modes is in principle possible using space-based platforms.  We now examine the
implications of our results for distinguishing between several competing theories for an optimistic
detection of $r=0.1$.  The following results are summarized in Table 2.

{\it{Inflation}}: General inflation predicts a tensor amplitude $P_h \propto H^2$ at lowest order in
slow roll.  The dominant energy condition demands that $\dot{H} < 0$ in a Friedmann Universe, leading
to a red tilted tensor spectrum, $n_T < 0$.  
The detection thresholds for the different experiments are presented in Table 1: while large tilts
are needed for Planck and ground-based experiments to make a decisive detection in the most
optimistic case, these missions are nonetheless capable of providing evidence in
support of inflation.  Future
missions will improve on these capabilities, but will still require relatively large tilts in the
most pessimistic outcome of small $r$.  Single field inflation predicts a consistency relation, $r = -8n_T$, shown
as a dashed curve in Figures 4 and 5.  Our results indicate that current probes and proposed
missions like CMBPol will fail to provide strong evidence in support of single field inflation.
If single field inflation is the true source of the B-mode, an ideal experiment will provide decisive
supporting evidence if $n_T < -0.02$, corresponding to $r > 0.15$ (c.f. Figure 4).  
\begin{figure}
\centering
\includegraphics[clip,width=3.25in]{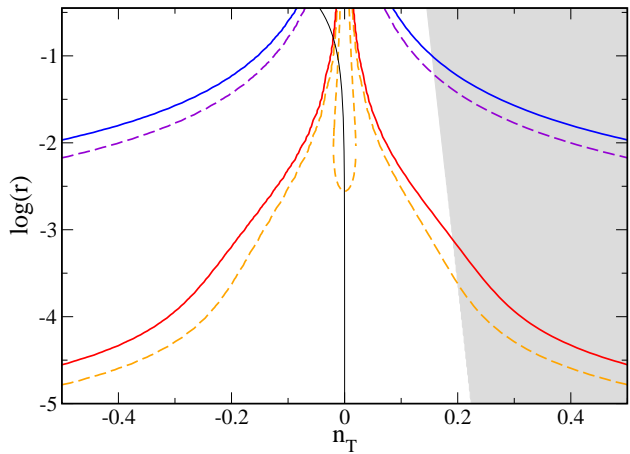}
\caption{Bayes factor as a function of $r$ and $n_T$ for CMBPol and the ideal CMB experiment. For CMBPol, thresholds for strong ($\ln B_{01} =-2.5$) and
decisive ($\ln B_{01} = -5$) evidence in favor of $\mathcal{H}_1$ are indicated by the top two contours: violet dashed and blue
solid curves,
respectively.  For the ideal mission, these same thresholds are given by the lower two contours: orange dashed and red solid curves,
respectively, and the central closed orange dashed contour indicates the threshold for strong evidence ($\ln B_{01} = 2.5$)
favoring the null hypothesis, $\mathcal{H}_0$.  The interpretation is that inside
this contour $n_T = 0$ is strongly favored over $n_T \neq 0$.  The gray region is excluded by
nucleosynthesis constraints, and the vertically-oriented solid black line gives the prediction of single field slow roll
inflation.}                                                                              
\end{figure}  

{\it String Gases}: During the quasi-static Hagedorn phase of string gas cosmology, the Hubble radius
shrinks so that fluctuation modes come to exist on cosmological scales \citep{Nayeri:2005ck}.  The anisotropic pressure of the matter
fluctuations generates large scale tensor perturbations; this pressure is smaller deep in the Hagedorn
phase than it is during the subsequent radiation dominated expansion \citep{Brandenberger:2006xi}.  The larger scale tensor modes,
which exit the horizon earlier, will therefore have smaller amplitudes than those that exit later, and
the
resulting tensor spectrum is a power law with a blue tilt.  Because it is a power law, $n_T$ is
bound by nucleosynthesis constraints, giving a small window within which $n_T$ is large enough to be
detected, but smaller than the nucleosynthesis cutoff.  CMBPol and the ideal satellite will provide
strong observational support if $0.12 < n_T < 0.15$ and $0.025<n_T<0.15$, respectively. 

{\it Loop Quantum Cosmology}: Motivated by loop quantum gravity, loop quantum cosmology (LQC)
is characterized by a quantum bounce joining a prior contracting phase to the expanding phase of big bang
cosmology.  It was discovered early that inverse volume corrections lead to a period of
superinflation ($w<-1$) in the early Universe \citep{Bojowald:2002nz}, generating a power law spectrum of tensor modes with an
unacceptably blue tilt \citep{Calcagni:2008ig}.  More recent analyses have studied the full
evolution of the perturbations from their generation during the contracting phase, across the
bounce, and through a period of slow roll inflation driven by a massive scalar field
\citep{Mielczarek:2009zw,Mielczarek:2010bh}.  Due to the shrinking Hubble radius, modes initiated during the contraction are strongly
blue, $P_h \sim k^2$.  After the bounce, those modes that are outside the horizon are
frozen and the blue spectrum is retained on these scales; modes within the horizon after the
bounce evolve during the subsequent period of inflation leading to a nearly flat tensor spectrum on
smaller scales.
The resulting spectrum is effectively a power law with a steep blue spectrum, $n_T \approx 2$, on
CMB scales, but the non-power law nature of the spectrum allows it to evade the nucleosynthesis
constraints.  This tensor tilt is large enough to be decisively detected by all the experiments
considered in this analysis. 
\begin{table*}
\caption{Conditions under which the different experiments will be able to provide strong supporting
evidence for different models of early Universe structure formation: single field inflation (SFI),
string gases (SG), loop quantum cosmology (LQC), and a nonsingular matter bounce (MB).  For LQC, we assume
that $n_T = 2$, large enough to be decisively detected by all experiments.  For the other experiments,
supporting evidence is obtained for the fiducial values of $n_T$ given.  These results are for $r=0.1$ for all models
except SFI, for which strong supporting evidence can only be found for $r>0.15$.  See the text for details.}  
\begin{tabular}{|l|c|c|c|c|}
  \hline
  Experiment & SFI & SG &LQC & MB\\ 
\hline
Planck & No &  No & Yes &No\\
\hspace{2mm} +PolarBear & No & No& Yes &No\\
\hspace{2mm} +QUIET (II)& No &  No& Yes & No\\
CMBPol&No& $0.12<n_T<0.15$&Yes&No\\
Ideal&$n_T<-0.02$&$0.025<n_T<0.15$&Yes&Yes\\
\end{tabular}
\end{table*}

{\it Matter Bounce}: String cosmology can also accommodate a bouncing Universe.  The early
implementations, Pre-Big Bang cosmology and the ekpyrotic scenario, included a singular bounce
connecting the contracting and expanding phases.  Due to the dynamics of the contracting phase, the
resulting tensor perturbations in the expanding phase are unacceptably blue in both models, with the
requirement that the tensor spectrum be suppressed.  Initially introduced as a toy model exhibiting a
nonsingular transition, a generic bounce in which the Universe passes from matter to radiation
domination prior to the bounce is capable of producing a scale invariant tensor spectrum
\citep{Finelli:2001sr}.
Perturbations generated during a collapsing dust dominated Friedmann Universe are identical to those
generated during de Sitter expansion \citep{Wands:1998yp}, since the amplitudes of all modes -- both sub- and
super-horizon -- grow at the same rate during dust dominated contraction.  This model is our null
hypothesis, $\mathcal{H}_0$, and  
it is phenomenologically distinct from all of the other models.  If $n_T = 0$, an ideal satellite will
provide strong evidence for it.

Although we have not considered them in this work, it is expected that
future space-based laser interferometers like Big Bang Observer and
Japan's Deci-hertz Interferometer Gravitational Wave Observatory (DECIGO) will measure $n_T$ at least as precisely as
an ideal CMB satellite \citep{Seto:2005qy,Kudoh:2005as,Zhao:2009mj} and should perform well under Bayesian model selection; we leave this analysis for future work.   
In the meantime, we can conclude that the awesome prospect of using the primordial B-mode as a window into the origin of
structure formation will be within the grasp of next-generation space probes.



\label{lastpage}

\end{document}